\documentclass[12pt]{iopart}
\begin{document}
\input epsf
\title{A study of nuclei of astrophysical interest in the continuum shell
model}
\author{K Bennaceur\dag, F Nowacki\dag\P,  J Oko{\l}owicz\dag\S ~and 
M P{\l}oszajczak\dag\footnote{Invited talk at the International 
Workshop on Physics with Radioactive Nuclear
Beams, \\ ~January 12--17, 1998, Puri, India}}
\address{\dag\ Grand Acc\'{e}l\'{e}rateur National d'Ions Lourds (GANIL), 
CEA/DSM -- CNRS/IN2P3, BP 5027, F-14076 Caen Cedex 05, France}
\address{\P\ Laboratoire de Physique Th\'{e}orique  Strasbourg (EP 106), 
3-5 rue de l'Universite, F-67084 Strasbourg Cedex, France }
\address{\S\ Institute of Nuclear Physics, Radzikowskiego 152, 
PL - 31342 Krakow, Poland}

\begin{abstract}
We present here the first
application of realistic shell model (SM) 
including coupling between many-particle
(quasi-)bound states and the continuum of one-particle scattering states to the
spectroscopy of $^{8}$B and to the calculation of
astrophysical factors in the reaction $^7\mbox{Be}(p,\gamma)^8\mbox{B}$.

\end{abstract}

\maketitle                                            

\section{Introduction}
The theoretical description of weakly bound exotic 
nuclei close to the drip-line is
one of the most exciting challenges today. What makes this subject both
particularly interesting and difficult, is the proximity of the particle
continuum implying strong modification of the effective nucleon--nucleon
interaction and causing unusual spatial properties of the nucleon density 
distribution (halo structures, large diffusivity). 
Many of those nuclei are 
involved in the chain of thermonuclear reactions and, in the absence of data   
at relevant energies, the models of
stars rely to certain extent on calculated 
astrophysical factors (see Bahcall (1989))~.

In weakly bound exotic systems, 
the number of excited bound states or narrow resonances 
is small and, moreover, they couple strongly to the particle continuum. 
Hence, these systems should be described in 
the quantum open system formalism which does not artificially separate
the subspaces of (quasi-) bound (the $Q$-subspace) and scattering (the
$P$-subspace) states. For well bound nuclei
close to the $\beta$-stability line,  
microscopic description of states in the first subspace is given
by nuclear SM with model-space dependent 
effective two-body interactions, whereas the latter subspace
is treated in terms of coupled channels equations. In this work, 
we question the validity of this
basic paradigm of nuclear physics, and propose
its modification for weakly stable exotic nuclei
by taking into account  
coupling between $Q$~ and $P$~ subspaces
in terms of residual nucleon--nucleon interaction.  This coupling
modifies the scattering solutions as well as the spectroscopic quantities for
interior bound states.

As said before, we are interested in
describing low lying bound and quasi-bound states in exotic nuclei. 
For that reason, we can restrict description of particle continuum to the
subset of one nucleon decay channels. Still, in few rare cases of
two-nucleon halo nuclei,
this limitation may turn out to be restrictive.
In any case, further improvement
of our model to more complicated channels like, e.g., $\alpha$ -
channels, can be done as well (Balashov \etal (1964)). 

\section{The shell model embedded in the continuum (SMEC)}
The influence of scattering continuum on the 
mean-field properties has been addressed before 
(see Dobaczewski \etal (1996) for review). No such
analysis has been done so far for the realistic SM. 
A possible starting point 
%of such a discussion
could be the Continuum Shell Model (CSM) approach 
(Fano 1961, Mahaux and Weidenm\H{u}ller (1969), Philpott (1977)), which in the 
restricted space of configurations generated using the finite-depth potential,
has been studied for the giant 
resonances and the radiative capture reactions 
probing the microscopic structure of these resonances (Barz \etal
(1977,1978), Fladt \etal (1988)). This is insufficient for nuclei 
close to drip lines, where it is essential to
have a most realistic description of bound state 
subspace. For that reason, the corner-stone  
of our approach, called the Shell Model Embedded in the Continuum (SMEC), is
the {\it realistic~SM} itself which is used
to generate the $A$-particle wavefunctions. This choice implies that 
the coupling between SM
states and the one-particle scattering continuum must be given by the residual
interaction. In our case, we use the residual interaction in the form :
\begin{eqnarray}
\label{force}
V = -V_0 (a + bP_{12}^{\sigma})\delta({\bf r}_1 - {\bf r}_2)  \ ,
\end{eqnarray}
with $a + b = 1$~ and $a = 0.73$~.

The key element of both SMEC and CSM is the treatment of 
single-particle resonances, which on one side 
may have an important amplitude inside a nucleus and, on other
side, they exhibit asymptotic behaviour of scattering wavefunctions
(Bartz \etal 1977).  The part of
resonance for $r < r_{c}$~, where $r_c$~ is the cut-off radius, 
is included in $Q$ subspace, 
whereas the remaining part is left in the $P$ subspace. The
wavefunctions of both subspaces are then renormalized 
in order to ensure the
orthogonality of wavefunctions in both subspaces.

In the SMEC calculations, we solve identical equations as in the CSM 
(Bartz \etal (1977)) but due to specificity of exotic
nuclei, ingredients of these calculations are modified. 
For the bound states we solve the 
SM problem : $H_{QQ}{\Phi}_i = E_i{\Phi}_i$~, 
using the code {\sc ANTOINE} (Caurier (1989)). $H_{QQ} \equiv QHQ$~ 
%which is the part of Hamiltonian acting in the $Q$ - subspace, 
is {\it identified} with the
realistic SM Hamiltonian and ${\Phi}_i$~ are the $A$-particle
(quasi-) bound wavefunctions. The quasi-bound resonances in the
continuum are included as well. 
\begin{figure}[t]
\begin{center}
%	\fbox{\psfig{figure=states_B8_0.eps,width=0.85\textwidth}}
%	\epsfxsize\hsize
	\epsfxsize0.8\hsize
	\epsffile{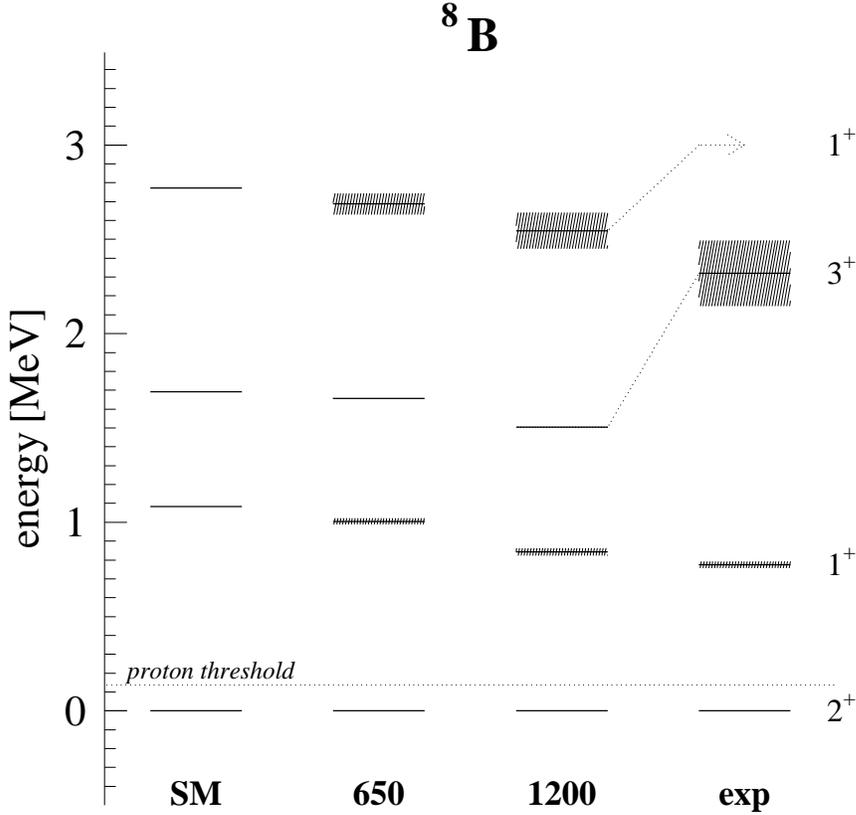}
\end{center}
	\caption{SM with Cohen--Kurath interaction and SMEC
	(labeled by residual interaction strength $V_0$) 
	vs.\ experimental $T=1$ states of $^8$B nucleus.  The proton threshold
	energy is adjusted to reproduce position of the ground state. The
shaded regions represent the width of resonance states.} 
	\label{b80f}
\end{figure}

For the continuum part, we solve the coupled channel equations : 
\begin{eqnarray}
\label{esp}
(E^{(+)} - H_{PP}){\xi}_{E}^{c(+)} = 0   \ ,
\end{eqnarray} 
where index $c$~ denotes 
different channels and $H_{PP} \equiv PHP$~. The sign $\pm$ characterizes the
boundary conditions, i.e., whether we consider incoming `$-$' or outgoing `$+$'
scattering waves. In our case, we have ingoing wave in the input channel 
and outgoing waves in all channels.  The structure of
$(A - 1)$~-~nucleus is given by the SM, whereas one nucleon
occupies a scattering state. The channel states are defined by coupling one
nucleon in the continuum to a `hole state' of $(A - 1)$~-~nucleus. 
The SM wavefunction has an 
incorrect asymptotic behaviour for states in the scattering continuum. On the
other hand, the standard approach consisting of adjusting values of 
the two-body matrix elements to the experimental data and/or 
modifying the monopole term of the effective interaction 
(Dufour and Zuker (1996)), 
makes the definition of average single-particle field and the radial
dependence of the single-particle wavefunctions somewhat
arbitrary. Therefore, to generate both
single-particle resonances and the radial formfactors of
occupied orbits entering the coupling matrix elements (\ref{force}) 
between states in the subspaces $Q$~ and $P$~, we use 
the finite-depth average potential of Saxon-Woods type with spin-orbit
part included. The parameters of the average potential
are fitted to reproduce experimental single-particle states, 
whenever their identification is possible. This is the initial choice of
potential, because the microscopic coupling of bound and scattering states 
generates the correction term to the mean-field which modifies
the single-particle wavefunctions for each $J^{\pi}$~ of the quasi-bound state 
. This correction term must be carefully taken
into account otherwise, the 
solutions including coupling of $Q$~ and $P$~ subspaces could be non-orthogonal.

The third system of equations are the coupled channel equations :
\begin{eqnarray}
\label{coup}
(E^{+} - H_{PP}){\omega}_{i}^{(+)} = H_{PQ}{\Phi}_i \equiv w_i   
\end{eqnarray}   
with the source term $w_i$~, which is given by the SM 
structure of $A$ - particle wavefunction for state ${\Phi}_i$~.
These equations define functions ${\omega}_{i}^{(+)}$~  which
describe the decay of quasi-bound state ${\Phi}_i$~ in the continuum. 
The source $w_i$~ couples the wavefunction of $A$ - nucleon 
localized states with $(A - 1)$ - nucleon localized states + one nucleon in the
continuum. The formfactor of the source term is given by the same average
potential as used in the $P$-space (\ref{esp}). The residual coupling 
(\ref{force}) is also identical.
The complete solution can be expressed by
means of three functions : ${\Phi}_i$~, ${\xi}_{E}^{c}$~ and ${\omega}_i$ :
\begin{eqnarray}
\label{eq2}
{\Psi}_{E}^{c} = {\xi}_{E}^{c} + \sum_{i,j}({\Phi}_i + {\omega}_i)(E -
H_{QQ}^{eff})^{-1} <{\Phi}_{j}\mid H_{QP} \mid{\xi}_{E}^{c}>  \ ,
\end{eqnarray}
where $H_{QQ}^{eff} = H_{QQ} + H_{QP}G_{P}^{(+)}H_{PQ}$~ is the effective
SM Hamiltonian including the coupling to the continuum and
$G_{P}^{(+)}$~ is the Green function for the motion of single particle in 
$P$~. Matrix $H_{QQ}^{eff}$ is
symmetric and complex. It can be diagonalized by the orthogonal transformation
: ${\Phi}_i \longrightarrow {\tilde {\Phi}_j} = {\sum}_{i}^{} b_{ji}{\Phi}_i$~
with complex eigenvalues ${\tilde {E_i}} - \frac{1}{2}i{\tilde {{\Gamma}_i} }$.
The eigenvalues of $H_{QQ}^{eff}$~ at energies $E = E_i$~ determine the
energies and widths of resonance states. With this changes, one obtains :
\begin{eqnarray}
\label{cons}
{\Psi}_{E}^{c} = {\xi}_{E}^{c} + \sum_{i}^{}{\tilde {\Omega}_i} 
[E - {\tilde E_i}
+ (i/2){\tilde {\Gamma}_i}]^{-1} <{\tilde {\Phi}_i} \mid H \mid {\xi}_{E}^{c}>
\ ,
\end{eqnarray}
for the wavefunction of continuum state modified by the discrete states and :
\begin{eqnarray}
\label{diss}
{\tilde {\Omega}_i} = {\tilde {\Phi}_i} + \sum_{c}
\int_{{\varepsilon}_c}^{\infty} dE^{'} {\xi}_{E^{'}}^{c} (E^{(+)} - E^{'})^{-1}
<{\xi}_{E^{'}}^{c}\mid H \mid {\tilde {\Phi}_i}>  \ ,
\end{eqnarray}
for the wavefunction of discrete state modified by the continuum states.

\section{Results}
\subsection{Spectroscopy of $^{8}B$ nucleus}
As an example of SMEC calculations for bound and resonant states, 
let us consider the proton-rich nucleus $^{8}$B~.
Fig.~1 compares SM spectrum for $T=1$~ states of $^{8}$B~ calculated
in the $p$ - shell using the
Cohen and Kurath (1965) (CK) interaction, 
with those obtained in the SMEC for two different 
strength $V_0$~ of the residual interaction (\ref{force}). In the SMEC
calculations, the Saxon-Woods average potential with
parameters : $U_0=-31.924$ MeV, $U_{so}=6.86$ MeV, $R_0=R_{so}=2.95$ fm and
$a=a_{so}=0.52$~ was used as a starting guess. The spectrum of
$^{8}$B does not show large sensitivity to the choice of these parameters. 
This initial average potential is then 
modified for each $J^{\pi}$~ state by the coupling to the
continuum. This correction make the average potential deeper, it increases
slightly the diffusivity and produceses a maximum at the center of the
potential. It is the corrected potential which is then used to calculate the
radial formfactors of coupling matrix elements 
and single-particle wavefunctions. 
The coupling to the continuum mixes the unperturbed SM
states. For example for $V_0=1200$ MeV${\cdot}$fm$^{3}$ in (\ref{force})
, the wavefunction for the first $1_{1}^{+}$ resonance 
has an overlap of 0.992, $0.1+0.005\,i$~ and $0.07+0.003\,i$~ 
with unperturbed SM states 
$1_{1}^{+}$, $1_{2}^{+}$ 
and $1_{3}^{+}$~, respectively. The calculated width of $1_{1}^{+}$
state is 17~keV and 29~keV for $V_0=650$ and 1200 MeV${\cdot}$fm$^{3}$~
respectively. The
experimental width for this state is $37 \pm 5$~keV. One should notice that also
the energy difference between $2_{1}^{+}$~ and $1_{1}^{+}$ states is 
improved by inclusion of the coupling to the continuum.

\subsection{Radiative capture reaction $^{7}Be(p,\gamma)^{8}B$~
at low energies}
The ${\beta}^{+}$~ decay of $^{8}$B, which is formed by the reaction 
$^{7}\mbox{Be}(p,\gamma)^{8}\mbox{B}$ at the center of mass (c.m.) energy
of about 20 keV, 
is the main source of high energy solar neutrinos. In the absence of data
in this region, the input of standard model for the solar 
neutrino problem (Bahcall and Ulrich (1988)) should be compared with those of
\begin{figure}[t]
\begin{center}
	\epsfxsize0.95\hsize
	\epsffile{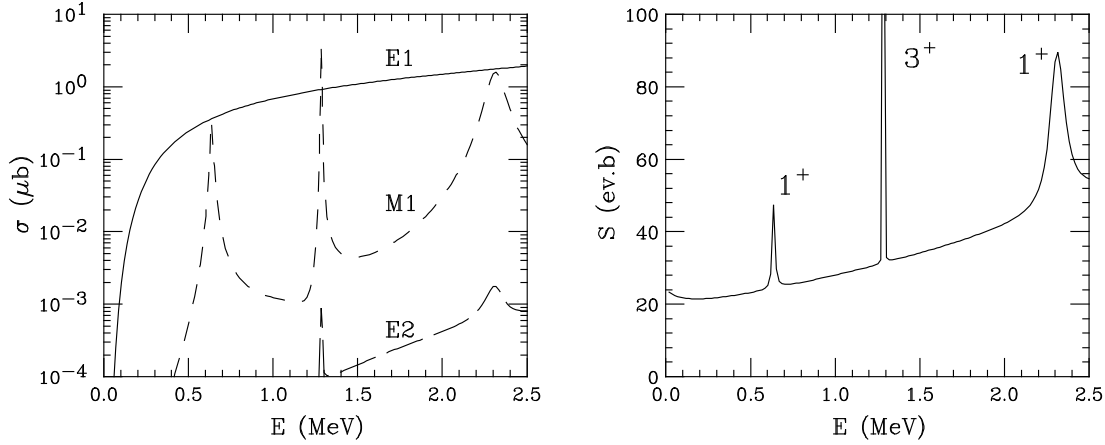}
\end{center}
	\caption{Multipole contributions to the total capture cross section
(left hand side) and the astrophysical $S$-factor (right hand side) 
of $^{7}\mbox{Be}(p,\gamma)^{8}$B as a function of the center of mass energy.
The SMEC calculations has been done with the residual interaction strength
$V_0=650$ MeV$\cdot$fm$^{3}$.}
\end{figure}        
various calculations. 
In the SMEC, the initial wavefunction 
${\Psi}_i([^{7}\mbox{Be} + p]_{J_i^{\pi}=1^{+}})$~ is : 
\begin{eqnarray}
\label{psiin}
\Psi_i(r)=\sum_{l_a j_a}i^{l_a}{\psi_{l_a j_a}^{J_i}(r)\over r}
\biggl[\bigl[Y^{l_a}\times\chi^{s}\bigr]^{j_a}\times\chi^{I_t}\biggr]^{(J_i)}
_{m_i}  
\end{eqnarray}
and the final wavefunction ${\Psi}_f([^{8}\mbox{B}]_{J_f^{\pi}=2^{+}})$ is:
\begin{eqnarray}
\label{psifin}
\Psi_f(r)=\sum_{l_b j_b}A_{l_bsj_b}^{j_bI_bJ_f} {u_{l_b j_b}^{J_f}(r)\over r}
\biggl[\bigl[Y^{l_b}\times\chi^{s}\bigr]^{j_b}\times\chi^{I_t}\biggr]^{(J_f)}
_{m_f}  \ .
\end{eqnarray}    
$I_t$~ and $s$~ denote the spin of target nucleus and  
incoming proton, respectively. 
$A_{l_bsj_b}^{j_bI_bJ_f}$~ is the coefficient of fractional parentage and
$u_{l_bj_b}^{J_f}$~ is the s.p. wavefunction in the many-particle state $J_f$~.
With the wavefunctions ${\Psi}_i(r)$~ and ${\Psi}_f(r)$~, 
we calculate the transition amplitudes:
\begin{eqnarray}
\label{x1}
\matrix{
\qquad T^{E{\cal L}} = C(E{\cal L})i^{l_a} \hat J_f \hat l_b \hat j_b\hat j_a
  <{\cal L} \delta J_f m_f \mid J_i m_i>
  <l_b 0 {\cal L} 0 \mid l_a 0> \hfill\cr
  \hfill\times W(j_b I_t {\cal L} J_i J_f j_a)
  W(l_b s {\cal L} j_a j_b l_a)
  I_{l_a j_a, l_b j_b}^{{\cal L},J_i} \qquad\cr}
\end{eqnarray}
for $E1$~ and $E2$~ and :
\begin{eqnarray}
\label{x2}
\matrix{
 T^{M1} = i^{l_a} \mu_N\hat J_f
  <1 \delta J_f m_f \mid J_i m_i> \hfill\cr
\qquad\times
\Biggl\{ W(j_b I_t 1 J_i;J_f j_a) \hat j_a \hat j_b \hfill\cr
  \hfill \Biggl[
    \mu\biggl({Z_t\over m_t^2}+{Z_a\over m_a^2}\biggr) 
    \hat l_a \tilde l_a
    W(l_b s 1 j_a;j_b l_a)
+ (-1)^{jb-ja} 2 \mu_a \hat s  \tilde s
    W(s l_b 1 j_a;j_b s)\Biggr] \cr
\hfill + \mu_t (-1)^{J_f-J_i} \hat I_t \tilde I_t
 W(I_t j_b 1 J_i J_f I_t) \delta_{j_a j_b}
\Biggr\} \delta_{l_a l_b} I_{l_a j_a, l_b j_b}^{0,J_i} \cr}
\end{eqnarray}
for $M1$ transitions, respectively. In the above formula,
$\delta=m_i-m_f,\ \hat a \equiv \sqrt{2a+1},
\ \tilde a \equiv \sqrt{a(a+1)}$ and
$I_{l_a j_a, l_b j_b}^{{\cal L},J_i} =\int u_{l_b j_b}
r^{\cal L}\psi_{l_a j_a}^{J_i} dr$~. The radiative capture cross section
can then be expressed in terms of those amplitudes as : 
\begin{eqnarray}
\label{tran}
\sigma^{E1,M1} = {16\pi\over9} \biggl({k_\gamma\over k_p}\biggr)^3
  \biggl({\mu\over\hbar c}\biggr)
  \biggl({e^{2}\over\hbar c}\biggr)
  {1\over 2s+1}~{1\over 2I_t+1}
\sum\mid T^{E1,M1}\mid^2  
\end{eqnarray}
\begin{eqnarray}
\label{tran1}
\sigma^{E2} = {4\pi\over75} \biggl({k_\gamma^5\over k_p^3}\biggr)
  \biggl({\mu\over\hbar c}\biggr)
  \biggl({e^{2}\over\hbar c}\biggr)
  {1\over 2s+1}~{1\over 2I_t+1}
\sum\mid T^{E2}\mid^2  
\end{eqnarray}
$\mu$~ stands for the reduced mass of the
system. 
 Fig.~2 shows the calculated multipole contributions to the total
capture cross section and the astrophysical $S$-factor as a function of the
c.m.\ energy. The calculation is done for $V_0 = 650$ MeV$\cdot$fm$^{3}$~. 
The proton threshold energy is adjusted to agree energies of calculated and
experimental $1^{+}$~ state. The photon energy is given by the difference of
c.m.\ energy of $[^{7}\mbox{Be} + p]_{J_{i}=1_{1}^{+}}$~ system 
and the experimental energy of the $2_{1}^{+}$~ ground state
of $^{8}$B~. The value of the
$S$-factor at c.m.\ energy of 20~keV: $S=22.58$ eV${\cdot}$b~, agrees with 
certain older experiments (Parker (1966), Kavanagh \etal (1969)) but
disagrees with the most recent one of Hammache \etal (1998). The ratios of SMEC
$S$ factors at different c.m.\ energies are: $S(20)/S(100)=1.07$ and
$S(20)/S(500)=1.05$. The ratio of $M1$ and $E1$ contribution is:
${\sigma}^{M1}/{\sigma}^{E1}=8.13{\cdot}10^{-6}$, $4.55{\cdot}10^{-5}$ and
$4.3{\cdot}10^{-3}$~ at 20, 100 and 500 keV, respectively. The
resonant part of $M1$~ transitions, which yields the contribution of
$S^{M1}=24.72$ eV${\cdot}$b at the $1_{1}^{+}$
resonance energy, decreases fast and becomes $S^{M1}= 9.6{\cdot}10^{-2}$,
$9.6{\cdot}10^{-4}$, $1.8{\cdot}10^{-4}$ eV${\cdot}$b 
at c.m.\ energies 500, 100, 20 keV, respectively.
This value is proportional to 
the square of spectroscopic amplitude of $p$-states, 
which for the CK interaction is 
$-0.352$ and $0.567$ for $p_{1/2}$ and $p_{3/2}$ respectively. Similar small 
values of spectroscopic amplitudes are
obtained for Kumar (1974) and PTBME \etal (1992) interactions (see Brown
\etal (1996)). At the
position of $1^{+}$ resonance, the predicted by SMEC calculations 
value of $S$-factor ($S = 47.05$ eV${\cdot}$b) 
is smaller than seen in some experiments (Kavanagh \etal (1969), 
Vaughn \etal (1970), Filippone \etal (1983)). Part of this discrepancy could be
related to the absence of quenching factor in our calculations.

\begin{figure}[t]
%\begin{small}
\begin{center}
	\epsfxsize0.95\hsize
	\epsffile{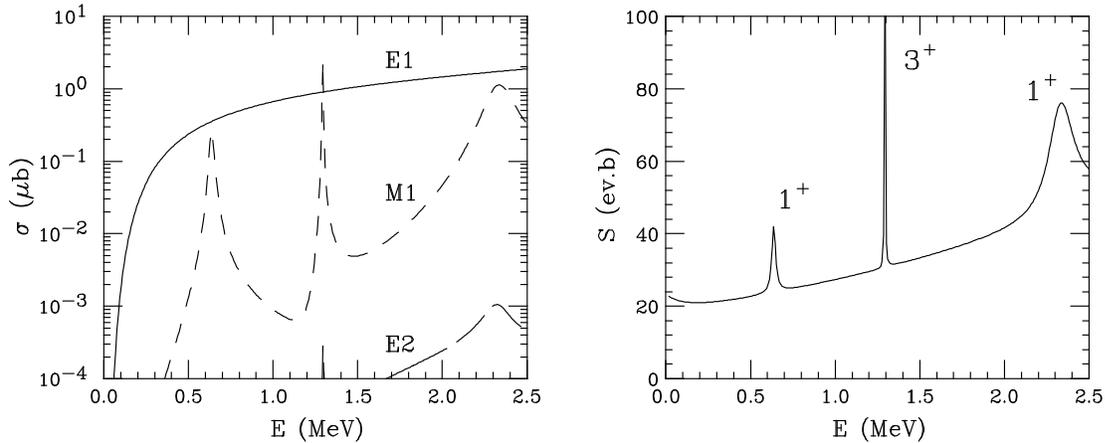}
\end{center}
	\caption{Multipole contributions to the total capture cross section
(left hand side) and the astrophysical $S$-factor (right hand side) 
of $^{7}\mbox{Be}(p,\gamma)^{8}$B as a function of the center of mass energy.
The SMEC calculations has been done with the residual interaction strength
$V_0=1200$ MeV$\cdot$fm$^{3}$.}
\end{figure}        
Fig.~3 shows the same as Fig.~2 but for $V_0=1200$ MeV$\cdot$fm$^{3}$.
Now, the calculated value of the
$S$ factor at c.m.\ energy of 20 keV is $S=22.70$ eV${\cdot}$b, rather close to
the value obtained for $V_0=650$ MeV${\cdot}$fm$^{3}$, and
disagrees with the recent experimental value given by Hammache \etal (1998)~. 
One should
stress however, that different experiments show significant fluctuations in the
extrapolated values of $S$~.
The ratios of $S$ factors at different c.m.\ energies:
$S(20)/S(100)=1.07$ and $S(20)/S(500)=0.997$, show much stronger
dependence on energy than seen for $V_0=650$ MeV${\cdot}$fm$^{3}$~. 
The resonant part of $M1$~ transitions at the $1_{1}^{+}$ resonance energy, 
yields a somewhat smaller value : $S^{M1}=18.23$ eV${\cdot}$b, than for
$V_0=650$ MeV${\cdot}$fm$^{3}$~. Finally, at the
position of $1^{+}$ resonance, the predicted by SMEC calculations 
value of $S$-factor ($S = 41.97$eV${\cdot}$b) 
is somewhat smaller than seen in the experiments and for $V_0=650$
MeV${\cdot}$fm$^{3}$.

\section{Conclusions}
In this work we have shown results of 
first calculations using the SMEC which couples the
realistic SM solutions for (quasi-) bound states with the scattering solutions
of one-particle continuum. The application to $^{7}\mbox{Be}(p,\gamma)^{8}$B
reaction yields satisfactory description of different components 
of the radiative capture cross section, including the resonant components.
In future, more unstable nuclei should be studied in the SMEC approach 
to systematically address the problem of effective interactions in the extreme
conditions of exotic nuclei.  At present, we are applying the SMEC  
to many other reactions of astrophysical interest such as
$^{14}\mbox{C}(n,\gamma)^{15}\mbox{C}$, $^{16}\mbox{O}(p,\gamma)^{17}\mbox{F}$,
$^{18}\mbox{O}(n,\gamma)^{19}\mbox{O}$.

\ack
Authors would especially like to thank S. Dro\.zd\.z and I. Rotter for many 
clarifying discussions. We are grateful to E. Caurier and A. Lefebvre 
for encouragement and useful discussions. The work was
partly supported by KBN Grant No. 2 P03 B 14010 and the Grant No. 6044
of the French - Polish Cooperation.
\vfill
\newpage

\section*{References}
\begin{harvard}
\item[]Bahcall J N, {\it Neutrino Astrophysics} (Cambridge
University Press, 1989).
\item[]Bahcall J N and Ulrich R K 1988 {\RMP} {\bf 60} 297 
\item[]Balashov V V \etal 1964 {\NP} {\bf 59} 414
\item[]Bartz H W \etal 1977 {\NP} {\bf A 275} 111
\item[]\dash 1977 {\NP} {\bf A 307} 285
\item[]Bertulani C A 1996 {\ZP} {\bf A 356} 293
\item[]Brown B A \etal 1996 {\NP} {\bf A 597} 66
\item[]Caurier E 1989 {\it unpublished}
\item[]Cohen S and Kurath D (1965) {\NP} {\bf A 73} 1
\item[]Dobaczewski J \etal 1996 {\PR} {\bf C 53} 2809
\item[]Dufour M and Zuker A P 1996 {\PR} {\bf C 54} 1641
\item[]Fano U 1961 {\PR} {\bf 124} 1866
\item[]Filippone B W \etal 1983 {\PRL} {\bf 50} 452
\item[]Fladt B \etal 1988 {\APNY} {\bf 184} 254, 300
\item[]Hammache F 1998 {\PRL} {\it in print}
\item[]Julies R E \etal 1992 {\it S. Afr. J. Phys.} {\bf 15} 35
\item[]Kavanagh W 1969 {\it Bull. Am. Phys. Soc.} {\bf 14} 1209
\item[]Kumar N 1974 {\NP} {\bf A 235} 221
\item[]Mahaux C and Weidenm\H{u}ller H 1969 {\it Shell-Model Approach to Nuclear
Reactions} (Amsterdam: North-Holland)
\item[]Parker P D 1966 {\PR} {\bf 150} 851
\item[]Philpott R J 1977 {\NP} {\bf A289} 109
\item[]Vaughn F J \etal 1970 {\PR} {\bf C 2} 1657
\end{harvard}

\end{document}